\documentclass[
    ,final            
  ]
  {aipproc}

\usepackage{axodraw}
\layoutstyle{6x9}

\begin{document}

\rightline{\small IPPP/09/74, DCPT/09/148}
\vskip -0.2in

\title{Signatures of Sneutrino NLSP in Gravitino Dark Matter Scenario at the LHC}

\classification{12.60.Jv, 14.80.Ly }

\keywords      {Supersymmetric models, supersymmetric partners of known particles}

\author{Yudi Santoso}{
  address={Institute for Particle Physics Phenomenology, Department of Physics, University of Durham, Durham, DH1 3LE, UK}
}

\begin{abstract}
We present the phenomenology of a supersymmetric model with sneutrino as the next to lightest supersymmetric particle (NLSP) in the gravitino dark matter scenario at the LHC. We focus on the leptonic signatures and study the feasibility of a particular trilepton signature as a possible discovery channel of supersymmetry.  
\end{abstract}

\maketitle


\section{Introduction}

The search for physics beyond the standard model has been a long endeavor in particle physics. Our hope is now on the Large Hadron Collider (LHC) experiment which is supposed to start getting physics data next year, in 2010. If there is a new physics up to $\sim 1$~TeV, such as predicted by most supersymmetric models, the LHC should be able to see it. To discover the new physics, however, it might not be sufficient to have a great machine and a huge amount of data, but we also need a correct theoretical interpretation to analyse the data. It is well known that the phenomenology of supersymmetry depends on its spectrum, therefore it is necessary to study all possible cases. The most studied scenario is with a neutralino as the lightest supersymmetric particle (LSP). Neutralino LSP is popular due to its possibility as the dark matter particle, provided that it is stable due to some parity conservation. Nevertheless, neutralino is not the only dark matter candidate within supersymmetry. It has been shown that gravitino, as the LSP, is also a feasible candidate~\cite{GDM}. In this scenario, the next to lightest supersymmetric particle (NLSP) is practically stable with respect to the collider detectors, while the gravitino itself is not detectable. The phenomenology, thus, depends on what the NLSP is. 

Here we focus our attention onto one such scenario, i.e. with sneutrino as the NLSP. As a neutral particle, the sneutrino would be seen as missing energy, just like the neutralino. However, because of the different mass hierarchy, the cross sections and the signatures are, in general, different.

\section{The Model}

The model that we study here is based on a supersymmetric model with non-universal Higgs masses (NUHM)~\cite{NUHM}. The supersymmetric parameters in NUHM are the universal sfermion masses $m_0$, gaugino masses $m_{1/2}$, trilinear masses $A_0$, all are set at the GUT scale, the Higgs vev ratio $\tan \beta$, the Higgs mixing parameter $\mu$ and the CP odd Higgs mass $m_A$. We choose a model with 
\begin{equation}
\tan \beta = 10, \ m_0 = 100~{\rm GeV}, \ m_{1/2} = 500~{\rm GeV}, \ A_0 = 0, \ \mu = 600~{\rm GeV}, \ m_A = 2~{\rm TeV}.
\end{equation}
For this set of parameters, the tau-sneutrino is the lightest non-gravitino supersymmetric particle~\cite{snuNLSP}. This can be seen as a result of the non-universal terms, which collectively known as the $S$-term, in the RGE. If $S$ is negative and large, such as in our case, the left-sleptons become lighter than the right-sleptons, and because of the split by the $D$-term the sneutrinos are lighter than the left-charged-sleptons. In this case, the tau-sneutrino is lighter than the electron/muon-sneutrino because of the Yukawa term in the RGE.

The physical spectrum for the lighter supersymmetric particles (or sparticles) is shown in Table~\ref{tab:a}. The heavier sleptons are about 460-482~GeV, the heavier gauginos are above 600~GeV, and the squarks are above 1~TeV, except for the lightest stop which is about 724~GeV. Note that a similar model, but within the gauge mediated framework, is also studied at a preliminary level in~\cite{CK}.


\begin{table}
\begin{tabular}{llr}
\hline
    \tablehead{1}{l}{b}{Particle} 
  & \tablehead{1}{r}{b}{ }
  & \tablehead{1}{r}{b}{Mass [GeV]}
   \\
\hline
tau-sneutrino & $m_{\tilde{\nu}_\tau}$ & 90.54  \\
lightest stau & $m_{\tilde{\tau}_1}$ & 115.31  \\
electron/muon-sneutrino & $m_{\tilde{\nu}_\ell}$ & 140.64  \\
$L$-selectron/smuon & $m_{\tilde{\ell}_L}$ & 161.42  \\
lightest neutralino & $m_{\tilde{\chi}^0_1}$ & 206.46  \\
lightest chargino & $m_{\tilde{\chi}^\pm_1}$ & 395.95  \\
second-lightest neutralino & $m_{\tilde{\chi}^0_2}$ & 396.10  \\
\hline
\end{tabular}
\caption{The lightest part of the supersymmetric spectrum in our model.}
\label{tab:a}
\end{table}

Note that in our model, the sneutrinos are the partners of the left-handed neutrinos. Left-sneutrino as dark matter is already excluded by direct dark matter detection experiments~\cite{snuDM}. However, assuming that gravitino is lighter than the sneutrino and is the dark matter, this model is still allowed.  For models including right-sneutrino effects in the RGE see~\cite{withsnuR}.

\section{The Collider Phenomenology}

Supersymmetric signatures at colliders come from the cascade decays of heavier sparticles down to the lightest stable sparticle. In our case, this is the sneutrino NLSP which appears as a large missing energy. The signatures can be classified as hadronic, pure leptonic or leptons plus jets. Eventhough the hadronic signals are expected to be stronger for a hadron collider such as the LHC, the heavy squark masses somewhat suppress the signals as compared to the standard model QCD background. A detail study is needed to find a promising signal in this category. Here we restrict ourselves to  the pure leptonic signatures instead. These come from the productions and decays of sparticles without strong interaction, i.e. the neutralinos, charginos, and sleptons.  

Since the neutralinos and charginos are relatively heavy, the trilepton signature through neutralino-chargino associated production, which could be prominent for the neutralino LSP scenario~\cite{CDF-trilep}, is suppressed in our model. On the other hand, the electron/muon-sneutrinos and the charged sleptons are relatively light. Therefore the dominant leptonic signals come from the production of these particles. However, for hadron colliders, such as the LHC, slepton pair production rates are not high, only around 100~fb for a 14~TeV LHC. Thus, the problem is to find signals that are strong enough  comparatively to the standard model (SM) backgrounds. The dominant SM backgrounds are from Drell-Yan, $W$ and $Z$.

\begin{figure}
\begin{picture}(220,132)(0,0)
  \DashLine(135,75)(165,95){2}  \Text(152,73)[]{$\tilde{\ell}^+$}
  \ArrowLine(165,95)(195,80) \Text(178,78)[]{$\tilde{\chi}^{+ \ast}$}
  \ArrowLine(165,95)(180,130)  \Text(172,130)[]{$\nu_\ell$}
  \DashLine(195,80)(220,105){2}  \Text(208,106)[]{$\tilde{\nu}_\tau$}
  \ArrowLine(195,80)(220,55)  \Text(207,57)[]{$\tau^+$}

  \Vertex(126,69){2}

  \DashLine(117,63)(87,43){2}  \Text(110,45)[]{$\tilde{\nu}_\ell$}
  \ArrowLine(87,43)(55,55) \Text(78,60)[]{$\tilde{\chi}^{+ \ast}$}
  \ArrowLine(87,43)(60,15)  \Text(70,9)[]{$\ell^-$}
  \DashLine(55,55)(25,85){2}  \Text(37,90)[]{$\tilde{\nu}_\tau$}
  \ArrowLine(55,55)(22,38)  \Text(35,35)[]{$\tau^+$}
\end{picture}
  \caption{Same sign taus trilepton signature through sneutrino-slepton associated production.} \label{fig:1}
\end{figure}
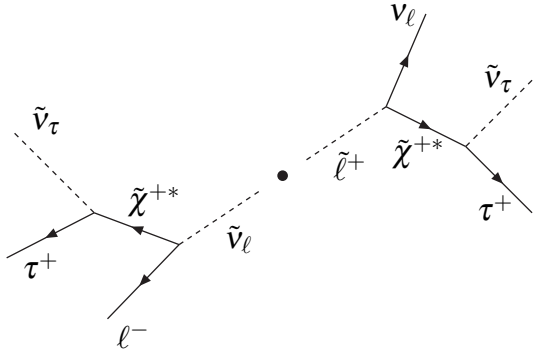

We studied 2, 3, 4 and 5-lepton plus missing energy signals. In general, we need a very large integrated luminosity to get $5\sigma$ discovery. Looking at the trilepton signals in more detail, the most dominant supersymmetric signal is from  $\tilde{\ell} \tilde{\nu}_{\ell}^\ast$ ($\tilde{\ell}^\ast \tilde{\nu}_\ell$) associated production, followed by $\tilde{\ell} \to \tilde{\nu}_\tau + \nu_\ell + \tau$ and $\tilde{\nu}_\ell \to \tilde{\nu}_\tau + \ell + \tau$ decays. For this channel we have two taus of the same sign and an electron or a muon of opposite sign as depicted in Fig.~\ref{fig:1}. The event rates are $\sim 10$~fb for negatively charged and $\sim 5.5$~fb for positively charged electron/muon (each), for a 14~TeV LHC. This, in principle, has a small SM background, coming only from $WWW$, as follows 
\begin{equation}
pp \to W^{+} W^{+} W^{-} \to \tau^{+} \nu_{\tau} \tau^{+} \nu_{\tau} \ell^{-} \bar{\nu}_{\ell}
\end{equation}
and similarly for the opposite signs. This background is of the order of $O(1)$~fb.
The supersymmetric background for this same sign taus signature is from neutralino-chargino ($\tilde{\chi}_2^0 \tilde{\chi}_1^\pm$),  but only about two order of magnitude lower. The supersymmetric QCD background should be reducible by counting the jet multiplicity. Imposing same-sign taus should also reduce the $t\bar{t}$ and fake leptons background. 


Nevertheless, the reality from the experimental point of view is not that simple. 
The difficulty is that tau decays very quickly such that it cannot be identified directly in the detectors. Instead, we have to rely on the reconstruction of tau from its decay products. Tau decays either leptonically ($\sim 35$\%) or hadronically ($\sim 65$\%). 
Thus instead of two taus and an electron or a muon, we get a signal with 0, 1 or 2 tau-jets (denoted by $\tau_h$). At this level, other backgrounds get in, most importantly from $WZ$ which can produce electrons and/or muons either directly or through taus. 
To utilize the same sign signature we take a look at the 2 $\tau_h$ case, $\tau_h^\pm \tau_h^\pm \ell^\mp$, where $\ell$ is either an electron or a muon. The $WZ$ background is through
\begin{equation}
pp \to W^+ Z \to \tau^+ \nu_\tau \tau^+ \tau^-  
\end{equation}
followed by the tau decays where the positively charged taus decay hadronically while the negatively charged tau decays leptonically. Our preliminary results show that eventhough this background has a relatively large event rate, the signal can still be distinguishable from the background because the lepton from the tau decay is softer. For example, by looking at the invariant mass distributions. Moreover, note that the $WWW$ background is also reduced by the tau decay branching ratios along with the signal, and therefore does not impose a further problem.

\section{Conclusion}

We have discussed, briefly, the phenomenology of a supersymmetric model with the tau-sneutrino as the effectively stable lightest sparticle at the LHC. A more complete and detailed analysis of this topic will appear in \cite{FRS}.
The analysis of this model illustrates the complexity in discovering beyond the standard model particle physics (BSM), i.e. not all models would be easily discovered (or excluded) at the LHC. A particular problem in our analysis is related to the fact that many taus are expected to appear in the signals due to the flavor of the lightest sneutrino. Thus, it is important for our model that taus can be identified correctly. Tau identification is often crucial for many BSMs. Lots of improvements, both on the experimental and the theoretical sides have been achieved on this issue~\cite{ATLAS-TDR,CMS-TDR}, so this should not be a big problem in the future. 
We are optimistic that technical difficulties, which cause delays at the LHC, could eventually be overcome, and we are hopeful that a new physics will be discovered in just a few years from now.


\begin{theacknowledgments}
I thank my collaborators on works related to this talk: John Ellis, Terrance Figy, Keith Olive and Krzysztof Rolbiecki; and Teruki Kamon for information on tau identification. 
\end{theacknowledgments}


\bibliographystyle{aipproc}   

\begin{thebibliography}{9}

\bibitem{GDM}
  J.~R.~Ellis, K.~A.~Olive, Y.~Santoso and V.~C.~Spanos,
  \emph{Phys.\ Lett.\  B} {\bf 588}, 7-16 (2004);
  J.~L.~Feng, S.~Su and F.~Takayama,
  \emph{Phys.\ Rev.\ D} {\bf 70}, 075019 (2004);
  \emph{Phys.\ Rev.\ D} {\bf 70}, 063514 (2004).


\bibitem{NUHM}
J.~R.~Ellis, K.~A.~Olive and Y.~Santoso,
  \emph{Phys.\ Lett.\ B} {\bf 539}, 107-118 (2002);
J.~R.~Ellis, T.~Falk, K.~A.~Olive and Y.~Santoso,
  \emph{Nucl.\ Phys.\ B} {\bf 652}, 259-347 (2003).

\bibitem{snuNLSP}
  J.~R.~Ellis, K.~A.~Olive and Y.~Santoso,
  \emph{JHEP} {\bf 0810}, 005 (2008).

\bibitem{CK}
  L.~Covi and S.~Kraml,
  \emph{JHEP} {\bf 0708}, 015 (2007).


\bibitem{snuDM}
  T.~Falk, K.~A.~Olive and M.~Srednicki,
  \emph{Phys.\ Lett.\  B} {\bf 339}, 248-251 (1994).

\bibitem{withsnuR}
  V.~Barger, D.~Marfatia and A.~Mustafayev,
  \emph{Phys.\ Lett.\  B} {\bf 665}, 242-251 (2008).

\bibitem{CDF-trilep}
  T.~Aaltonen {\it et al.}  [CDF Collaboration],
  \emph{Phys.\ Rev.\ Lett.}\  {\bf 101}, 251801 (2008).


\bibitem{FRS}
  T.~Figy, K.~Rolbiecki and Y.~Santoso,
  in preparation. 



\bibitem{ATLAS-TDR}
ATLAS Collaboration,  
  \emph{ATLAS detector and physics performance. Technical design report.}  Vol. 1 \& 2.
%

\bibitem{CMS-TDR}
  G.~L.~Bayatian {\it et al.}  [CMS Collaboration],
  \emph{J.\ Phys.\ G} {\bf 34}, 995-1579 (2007).


\end{thebibliography}

\end{document}